\documentclass[AMA,STIX1COL]{WileyNJD-v2}
\usepackage{bbm}
\usepackage{lscape}
\usepackage{subcaption}
\usepackage{xcolor}
\usepackage{booktabs}
\usepackage{makecell}
\newcommand{\supplementarysection}{%
  \setcounter{figure}{0}
  \let\oldthefigure\thefigure
  \renewcommand{\thefigure}{S\oldthefigure}
}

\articletype{Research Article}%

\newcommand{\N}{\mbox{Normal}}

\received{TBD}
\revised{TBD}
\accepted{TBD}

\begin{document}

\title{TITE-CLRM: Towards efficient time-to-event dose-escalation guidance of multi-cycle cancer therapies}

\author[1]{Lukas A. Widmer*}
\author[1]{Sebastian Weber}
\author[2]{Yunnan Xu}
\author[1]{Hans-Jochen Weber}

\authormark{WIDMER \textsc{et al}} %
\fundingInfo{Novartis Pharma AG, Basel, Switzerland and Novartis Pharmaceuticals, East Hanover, NJ, USA}

\address[1]{\orgdiv{Analytics}, \orgname{Novartis Pharma AG}, \orgaddress{\state{Basel}, \country{Switzerland}}}

\address[2]{\orgdiv{Analytics}, \orgname{Novartis Pharmaceuticals Corporation}, \orgaddress{\state{East Hanover, New Jersey}, \country{USA}}}

\corres{*~Lukas A. Widmer, Novartis Pharma AG, CH-4002 Basel, Switzerland. \email{lukas\_andreas.widmer@novartis.com}}

\abstract[Summary]{Treatment of cancer has rapidly evolved over time
  in quite dramatic ways, for example from chemotherapies, targeted
  therapies to immunotherapies and chimeric antigen receptor
  T-cells. Nonetheless, the basic design of early phase I trials in
  oncology still follows pre-dominantly a dose-escalation
  design. These trials monitor safety over the first treatment cycle
  in order to escalate the dose of the investigated drug. However,
  over time studying additional factors such as drug combinations
  and/or variation in the timing of dosing became important as
  well. Existing designs were continuously enhanced and expanded to
  account for increased trial complexity. With toxicities occurring at
  later stages beyond the first cycle and the need to treat patients
  over multiple cycles, the focus on the first treatment cycle only is
  becoming a limitation in nowadays multi-cycle treatment
  therapies. Here we introduce a multi-cycle time-to-event model (TITE-CLRM: Time-Interval-To-Event Complementary-Loglog Regression Model)
  allowing guidance of dose-escalation trials studying multi-cycle
  therapies. The challenge lies in balancing the need to monitor
  safety of longer treatment periods with the need to continuously
  enroll patients safely. The proposed multi-cycle time to event model
  is formulated as an extension to established concepts like the
  escalation with over dose control principle. The model is motivated
  from a current drug development project and evaluated in a
  simulation study. }


\jnlcitation{\cname{%
\author{LA Widmer}, 
\author{S Weber}, 
\author{Y Xu}, and 
\author{HJ Weber}} (\cyear{202x}), 
\ctitle{Towards efficient time-to-event dose-escalation guidance of multi-cycle cancer therapies}, \cjournal{Statistics in Medicine}, \cvol{202x;xx,x--xx}.}

\maketitle

\section{Dose-finding for compounds to treat cancer}\label{Introduction}
In clinical Phase 1 cancer trials, the dose finding generally follows
the maximum tolerated dose (MTD) concept and is assessed in the first
treatment cycle. The origin of this concept goes back to animal
toxicology studies which defined the MTD as:
\begin{quote}   
The MTD is defined as the highest dose of the test agent during the
chronic study that can be predicted not to alter the animals’
longevity from effects other than
carcinogenicity\cite{Sontag1976}.
\end{quote}

Cytotoxic agents to treat cancer came up in the 1960s. These compounds
induced high response rates and even cure. To overcome resistance,
intensive chemotherapies have been developed as combinations which are
given over a short period of time to avoid clonal escapes
\cite{Crawford2013, DeVita2008}. The MTD was of interest to achieve
this treatment objective acknowledging the narrow therapeutic window
of these therapies and the short treatment duration\cite{Luger2017}.

In oncology, the determination of the MTD became the leading concept
of dose finding over decades. This reflected the increasing efficacy
of cytotoxic chemotherapies with higher toxicity, as the mechanisms
responsible for both were intimately linked -- i.e., increasing
toxicity also caused increased efficacy, and identifying the MTD as
well as treating patients at the MTD was close to optimal. However,
this concept was later also followed for
compounds with a different mode of action that require longer
treatment duration than intensive chemotherapies.

Typically, the MTD is determined in dose escalation studies in which
doses are increased in small cohorts of patients until a predefined
threshold of toxicity is reached. Toxicity is assessed based on the
proportion of patients experiencing a dose limiting toxicity
(DLT) over the time period of the primary safety assessment
period. This primary safety assessment period is commonly just one
treatment cycle. However, the upcoming targeted therapies require continuous
therapy which has an impact on the dosing strategy: It requires longer
toleration by the patients, therefore, a single cycle observation
period is not sufficient especially in the presence of cumulative or
late-onset toxicities, which otherwise might lead to overdosing. This
is a common challenge which could even require post-approval dose
modification \cite{Shah2021}. A meta-analysis conducted in 2015
revealed that the suggested dose from drug labels of 62\% FDA-approved
compounds were lower than their MTD \cite{Sachs2016}. Inappropriate
doses can cause safety and tolerability issues, and further impact the
effectiveness of experimental therapies.

Extending the primary safety assessment period beyond one treatment
cycle raises several key challenges. For one, the trial escalation would
occur at a slower rate as more time would be needed for each patient
cohort to complete. At the same time, the issue of patient drop-out gets
amplified due to the longer time patients are being at risk. Many
established escalation designs disregard time and consider simply the
number of DLT events over the primary safety assessment
period. Whenever the primary safety assessment period is reasonably
short then the trial can proceed at a fast pace, which is not possible
with longer safety assessment period windows. 

Drop-out can occur by chance or is related to toxicity and thus, informative regarding DLT even if DLT criteria are not met. Ignoring patients who dropped out may impact the outcome (ICH E9 (R1)
addendum\cite{ICH2019}). A possible approach is the \textit{while-on-treatment strategy} where all information before the
drop-out is considered\cite{Mercier2024}. In designs with a fixed DLT observation period, data from patients who do not complete this period and have not experienced a DLT are typically discarded and patients are replaced. Such an approach may introduce selection bias in particular when the probability to drop out depends on the dose level. 

Allowing a trial to proceed with patient enrollment while using only
partially observed patient data is a critical consideration for a
practical dose escalation procedure. This has been recognized
previously in \cite{Cheung2000} which introduced the time-to-event
continual reassessment method (TITE-CRM). The TITE-CRM approach
extends the original continual reassessment method (CRM
\cite{quigleyCRM1990}) by weighting the model likelihood according to
the percentage of completed follow-up time. The more recently
introduced model-assisted approaches like the Bayesian optimal
interval design (BOIN \cite{Yuan2016}) have also been extended to allow for
partial follow-up of patients with the TITE-BOIN methodology
\cite{Yuan2018}. The approach relies on performing a single mean
imputation of the patient data with partial follow-up data. Here, we
pursue a different route by extending the two-parameter logistic CRM
introduced in \cite{Neuenschwander2008} (referred to as BLRM here) to
a particular form of a time-to-event model. The extension is done in
close alignment with key concepts of the BLRM methodology. As a
consequence the substantial experience with the BLRM can be carried
over to the new model. At the same time it allows us to take advantage
of standard time-to-event methodology, which addresses censoring and
drop-out in a coherent framework.

Turning to monitor multiple treatment cycles implicitly changes the
concept of the MTD concept, which is typically related to a single
cycle DLT observation period. As noted, for targeted therapies the
clinical main interest is to know the probability of a DLT over a time
interval exceeding just one cycle. We therefore refer to the following
definitions in this paper:
\begin{itemize}
    \item \textbf{by-cycle MTD}: The MTD determined based on a single cycle observation period
    \item \textbf{treatment-specific MTD}: The MTD determined on a therapy-specific DLT observation period which typically consists of multiple cycles. 
\end{itemize}

We illustrate the proposed approach using a case study in Acute
Myeloid Leukemia (AML) in Section \ref{Case_Study}. In this case
study, we incorporate prior information as well as information of
neighboring dose levels for decision making. We provide a detailed
statistical methodology in Section \ref{Methodology}. A simulation
study which corresponds to the setup of the case study presented in
Section \ref{Case_Study} is used to evaluate the performance of the
1-cycle and 3-cycle observation period time-to-(first)DLT and BLRM
models considering constant, de- and increasing probability of
toxicity profiles over time with and without informative and
non-informative dropout in Section \ref{Results}. We conclude with a
discussion of the proposed approach in Section \ref{Discussion}.

\section{Case study in post-transplant AML}\label{Case_Study}
\begin{figure}[h]
\centerline{\includegraphics[width=0.9\textwidth]{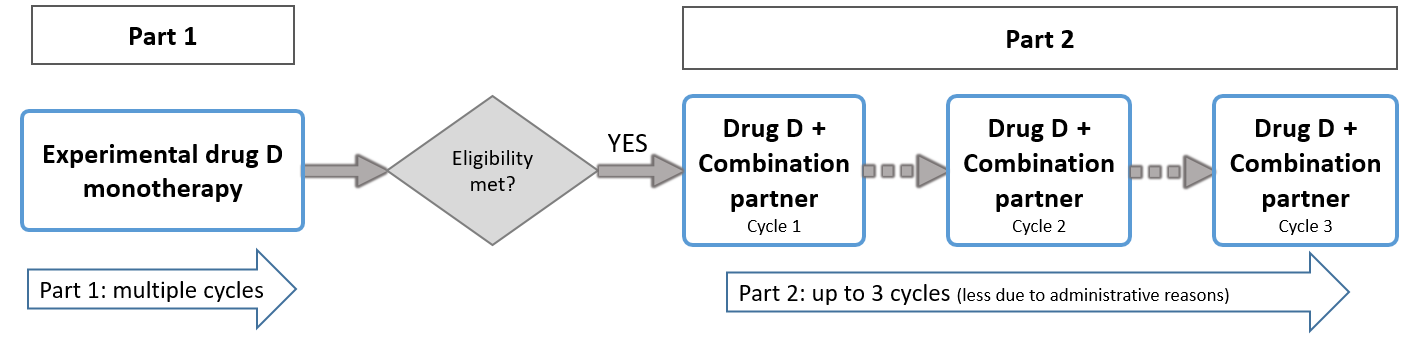}}
\caption{\label{Case study design}Case study design.}
\end{figure}
A single arm open-label multicenter study aims to find the recommended
dose levels of experimental \textit{drug D} as monotherapy and
subsequently in combination with a combination partner, \textit{drug
  C}, with late-onset toxicity in patients with AML. The study
consists of two sequential parts. Intrapatient escalation is
prohibited throughout the study. In Part 1, patients receive
\textit{drug D} monotherapy for multiple cycles. Subsequently, only
eligible patients will enter Part 2 and receive the combination of
\textit{drug D} and \textit{drug C} (Figure \ref{Case study
  design}). In both parts, different dose levels of \textit{drug D}
are available and the recommended dose levels of \textit{drug D} may
differ due to the addition of \textit{drug C} in Part 2.  Part 2
allows flexible numbers of cycles at a maximum of 3 cycles and
\textit{drug C} has a fixed ramp-up dosing schedule. With this set-up,
patients could enter Part 2 around the same time at different dose
levels of \textit{drug D}, and some patients may only undergo 1 cycle
while others may undergo 3 cycles. Therefore, traditional methods that
use a fixed duration of observation and a fixed dose level cohort may
not be appropriate for Part 2 dose finding.

With the proposed approach, a safety assessment will be performed on
the DLT data throughout Part 2 combination therapy. All patients will
be evaluated with the model upon completion of Cycle 1 in Part
2. Ideally, every time a new patient completes Cycle 1, the model will
be updated with DLT data from completed cycles of other patients
already in Part 2 in addition to the Cycle 1 DLT data of the new
patient. Hence, the model utilizes data from multiple cycles not
limiting to one single cycle.
 
The proposed approach does not require pause for the assessments and
enables operational ease. In case of intercurrent events such as
treatment discontinuations in Part 2, the safety data by the end of
the previous cycle(s) will be used, meaning that patients with such
intercurrent events will be kept in the analysis set rather than
replaced with additional enrolled patients.

\section{Methodology}\label{Methodology}

Established model-based methods like the one-parameter continual
reassessment model (CRM) or the two-parameter Bayesian logistic
regression model (BLRM) use a binomial likelihood to model the
probability $\pi$ for a DLT over the first cycle as a function of dose
$d$. These approaches use a Bayesian framework to obtain the posterior
probability $P(\pi(d)|y)$ for a DLT event at a set of prespecified
dose levels (with $y$ denoting the observed data), inferring the
dose-toxicity relationship. After each cohort of patients completing
cycle 1, the posterior is then used to determine a set of eligible
doses for the next cohort. An often used selection principle is the
escalation with overdose control (EWOC\cite{Babb1998}) criterion. EWOC requires that
the posterior probability $P(\pi(d)|y)$ for any given dose $d$ must
not exceed a limiting toxicity $\pi_c$ by some probability $p_c$ ,
$P(\pi(d) > \pi_c|y) < p_c$. Often $\pi_c$ of 33\% and $p_c$ of 25\%
is used, which is equivalent to requiring that the 75\% quantile of
the posterior for a DLT event at dose $d$, $P(\pi(d)|y)$, must be
smaller than 33\%. As these concepts are widely used and proven to
enable safe and efficient trials, we will generalize these accordingly
to the setting of multiple cycles.

\subsection{Time-to-DLT combination drug model likelihood}

\begin{figure}[h]
\centerline{\includegraphics[width=1.0\textwidth]{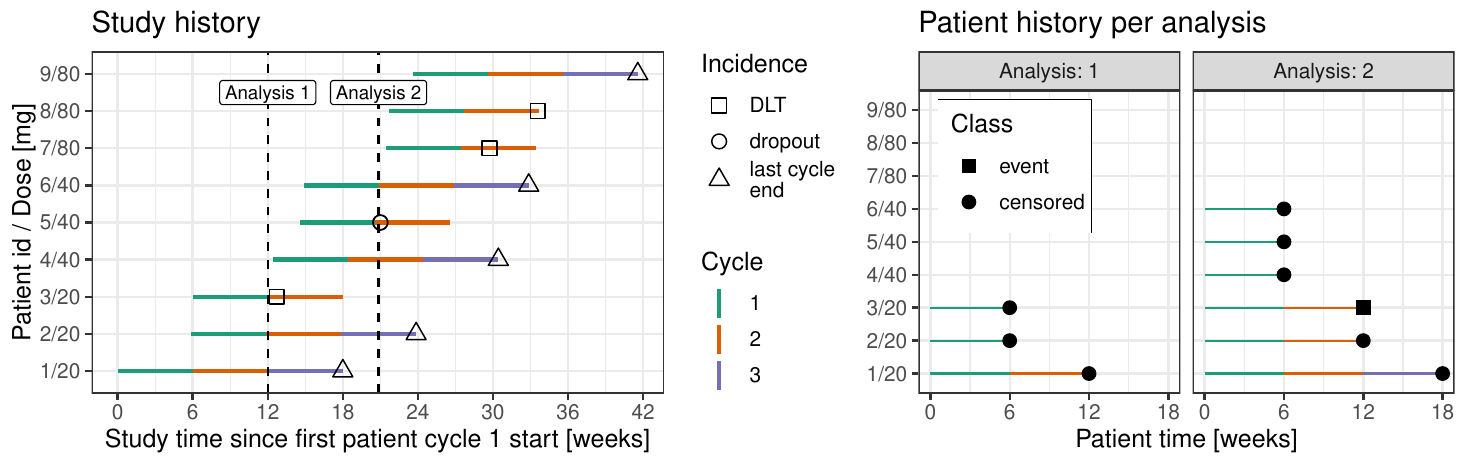}}
\caption{Trial event history in study time and trial event history in
  patient time at two analysis time points.\label{study_time}}
\end{figure}

In Figure \ref{study_time}, several patient journeys in a dose
escalation study are depicted. Patients start treatment and are
followed until the first cutoff of the first analysis time
point. Ideally we can use information documented up to this time point regardless if patients have reached the end of the scheduled
observation period. E.g. for patient 1/20 we can conclude that no DLT was observed during the first two treatment cycles. For patients 2/20 and 3/20 we know that no DLT occurred during Cycle 1. Patient 4/40 and later patients have not completed cycle 1 at the analysis cut-off and are not part of Analysis 1. So Analysis 1 includes 3 patients, and we consider patients 1-3 as censored.

Analysis 2 includes 6 patients. Further information is now available
for patients 1/20, 2/20 and 3/20. Patients 4/40, 5/40 and 6/40 are new patients for this analysis. Patients 1/20 and 2/20 are now considered censored at the end of cycle 2 and 3, respectively, and patient 3/20 is considered with a DLT (event) in cycle 2. Furthermore, patients 4/40 through 6/40 as censored after cycle 1. A sequential analysis makes efficient use of all available information, meaning completed
patient-cycles.

To account for longer time intervals due to multiple cycles and the
need for analysing partially observed patients who have not yet
completed all cycles but only a few as illustrated in Figure
\ref{study_time}, we use a time to first event model. For a cycle $j$, the probability for a DLT event is modeled as a function of the dose level $d_{j,k}$ ($k$ labeling the different dose levels) through a piece-wise constant time to first event model. The pieces are aligned with the duration of each cycle and the hazard for an event is taken to be constant throughout the cycle. Right censoring can occur for each patient and is assumed to be non-informative as random or administrative. Thus, for a patient $i$ one observes either the event time $T_i$ or the censoring time $C_i$, whatever occurs first. This is denoted as
\begin{eqnarray*}
  (U_i,\delta_i) & \mbox{ with } U_i = \min(T_i,C_i) \mbox{ and }
                   \delta_i = \mathbbm{1}{(T_i \leq C_i)},
\end{eqnarray*}
where $U_i$ represents the observed time, $\delta_i$ equals $1$ when an
event occurs and $\delta_i$ equals $0$ when no event occurs. As the
primary focus of the model is the risk for DLT events over the course
of a sequence of cycles to estimate the \textit{treatment-specific MTD}, we do not aim to model accurately the risk
for an event during the time-course of a cycle. One could model the
event and censoring times as interval censored observations, but we
disregard here the interval censoring and use a continuous time
representation of the time to event process. The continuous time
representation can be understood as an approximation of the interval
censored process with time lapsing in full units of a cycle. As a
consequence, the observed event time $T_i$ and censoring times $C_i$
are always recorded at the planned cycle completion time-point (time
lapses in units of cycles). Here we assume for simplicity that each
cycle $j$ has the same duration $\Delta$ such that any time being
recorded is a multiple of $\Delta$ and we denote time $t$ being during
a cycle $j$ with $t \in ((j-1) \, \Delta, j \, \Delta]$ or
  equivalently $t \in I_j$ for brevity.

Since DLTs are potentially caused either by the given experimental
drug or the background treatment used as standard of care, both
contribute to the overall DLT event process. We therefore use a
compound Poisson process such that the rate of DLT events is denoted
by the intensity function, which is for a compound Poisson process the algebraic sum of two intensity functions, with $h_1(t \in I_j)$ and $h_2(t \in I_j)$ representing the event process of the experimental drug and background treatment in cycle $j$, respectively. As the rate of DLT events may change over time we allow for changes in the intensity function between cycles such that the time to event model is a non-homogeneous compound Poisson process. Therefore, the intensity function is modelled per cycle $j$ for up to $J$ cycles defined by the maximal follow-up per patient. The intensity functions of each contributing process is modeled on the logarithmic scale to ensure their positivity.

\begin{eqnarray*}
 \text{Compound intensity function: } h(t \in I_j) &=& h_1(t \in I_j)  + h_2(t \in I_j)  \\
 \text{Experimental drug: }\log(h_1(t \in I_j)) &=& \log(h_{1,j}) = \alpha_1 + \beta_1 \,
  \log(d_{j,k}/\tilde{d}) \\
 \text{Background treatment: }  \log(h_2(t \in I_j)) &=& \log(h_{2,j})
 = \alpha_2 + (J-1) \, \gamma_2 \, \sum_{l=1}^{j-1} \xi_{2,l} \text{ ,
   with the constraint of}
\sum_{l=1}^{J-1} \xi_{2,l} = 1
\end{eqnarray*}

The intercept $\alpha_1$ determines the hazard for an event during the first cycle at the reference dose $\tilde{d}$ in absence of the
background treatment. The slope parameter $\beta_1$ is constrained to
be positive to ensure a monotonic increase in the hazard with
increasing dose. The background treatment is not dose dependent, but
its event intensity is allowed to increase monotonically over the
cycles. Thus, the cycle is considered as an ordinal variable
accounting for accumulation effects due to elongated treatment, which
can lead to an increase or decrease of the risk for a DLT in relation
to previous cycles as the sign of $\gamma_2$ is not constrained. This
monotonic modeling of ordinal variables is described in Bürkner
(2020)\cite{Buerkner2020}. Briefly, the $\gamma_2$ parameter is the average increase of the hazard between cycles and the sum over the
$\xi_2$'s is the fraction of the maximal effect due to the last cycle
as the $\xi_2$'s are constrained to sum to one.

Using basic event history analysis, the key quantities to characterize
the event process are

\begin{eqnarray*}
  H(t \in I_j) = H_j &= \Delta \sum_{l=1}^{i\leq j} h_l & \mbox{cumulative hazard} \\
  S(t \in I_j) &= \exp(- H_j ) & \mbox{survivor function, } P(T > t = j \, \Delta)
  \\
  f(t \in I_j) &= h_j \, \exp( -H_j ) & \mbox{probability density for an event
                                in cycle } j .
\end{eqnarray*}

The likelihood for all patients is then

\begin{equation}
  L(U|\alpha_1, \beta_1, \alpha_2, \gamma_2, \xi_2) = \prod_{i} f(T_i)^{\delta_i} \,
  S(C_i)^{1-\delta_i} .
\end{equation}

The specified model for the drug related event process is structurally closely related to the two-parameter Bayesian logistic regression model (BLRM) in Neuenschwander~(2008)\cite{Neuenschwander2008}. Discarding the background treatment, setting the cycle length to be unity and
considering that the probability for events (meaning not zero events)
during the first cycle is $\pi = 1-S(t=I_1)$, it follows that the
complementary log-log ($\mbox{cloglog}(x)=\log(-\log(1-x))$) of $\pi$
is equal to $\log(h_1)$. Hence, the considered model is a variant of
the BLRM in that it can be seen as a binomial model with a
complementary log-log instead of a logit link when restricted to the
first cycle only.

\subsection{Dose recommendation}

The EWOC principle determines the set of eligible doses whenever only
the first cycle is monitored for safety events. It therefore needs to
be generalized for the multi-cycle time to event setting. As it is a
clinical consideration whether one aims to control toxicity over (any) cycle or over the entire multi-cycle therapy, we propose two
alternative variants of the EWOC principle. Whenever one seeks to
control toxicity over the multi-cycle therapy, one may simply apply
the EWOC principle to the cumulative probability for a toxicity event
up to the end of follow-up with cycle $J$:
\begin{equation}
  Pr[P(t \in \{I_1, \dots, I_J\}| d) > \pi_c] < p_c .
\end{equation}
In case of a constant hazard for a toxicity event over all cycles this leads to a more
conservative dose recommendation whenever the commonly used thresholds of EWOC are used without adjustments to the longer exposure time of multiple cycles. In case the clinical focus is on controlling the toxicity for each cycle, we propose to control the probability for a toxicity event during any given cycle conditioned on no toxicity event in previous cycles. That is, the EWOC condition is only met if for all cycles the conditional probability for a toxicity event remains below the pre-defined threshold as
\begin{equation}
  \forall_{j=1}^J Pr[P(t \in I_j|t > (j-1) \, \Delta, d) > \pi_c] < p_c .
\end{equation}
Conditioning EWOC for the $j^\mathrm{th}$ cycle on no event in previous
cycles ensures that the formerly often used thresholds of $\pi_c$
(33\%) and $p_c$ (25\%) keep their original meaning, since the
exposure time of one cycle does not change. However, due to
accumulating toxicity effects, the multi-cycle EWOC may be more
stringent in practice. In case it is clinically meaningful to vary the admissible risks over the cycles, then the used thresholds can be
varied per cycle accordingly. Here we set the thresholds $\pi_c$ and
$p_c$ to be equal for all cycles.

\begin{figure}[htb]
\begin{subfigure}{\textwidth}
\caption{\label{fig:prior_summary_conditional}}
\centerline{\includegraphics[width=0.99\textwidth]{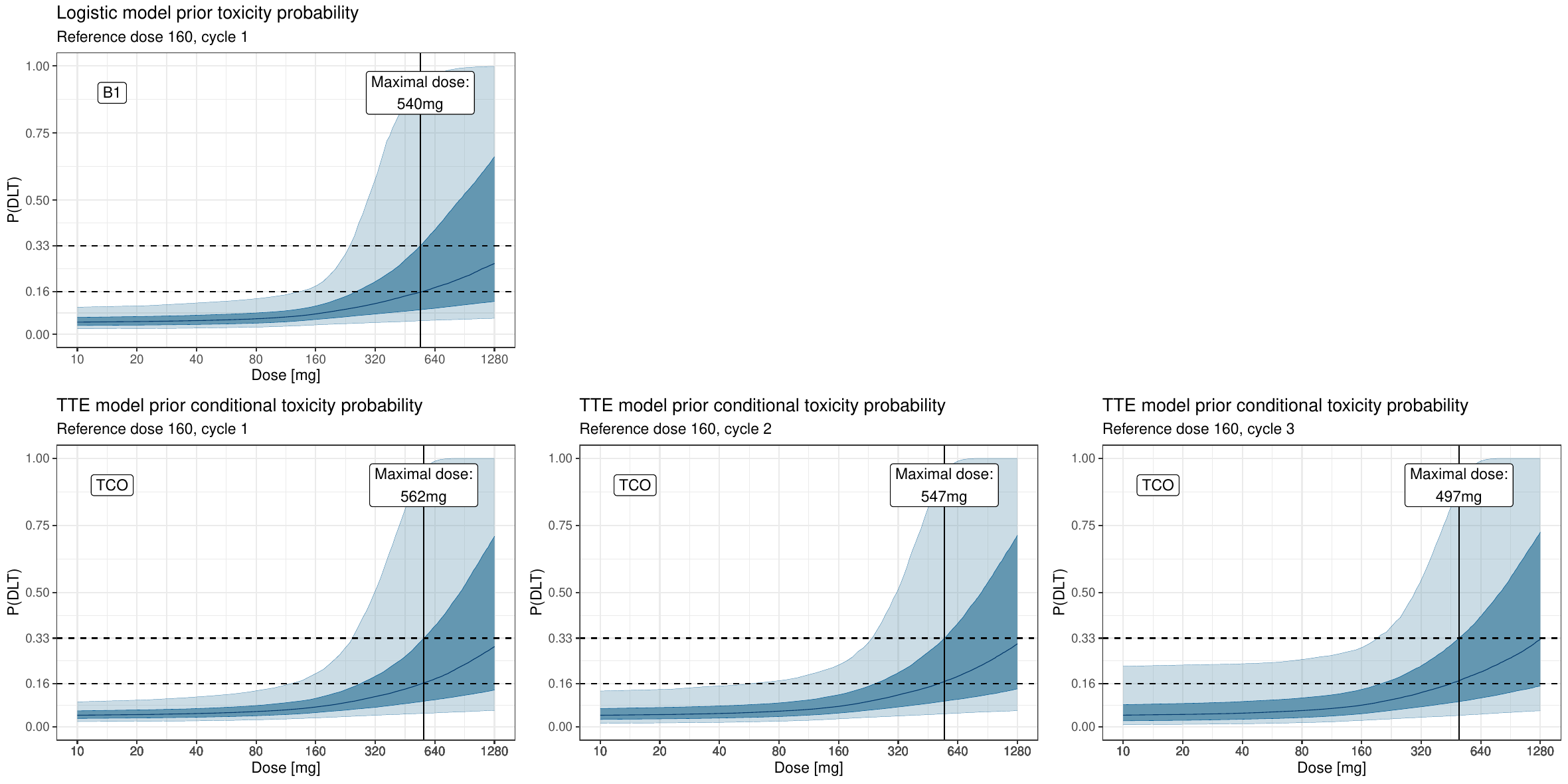}}   
\end{subfigure}
\begin{subfigure}{\textwidth}
\caption{\label{fig:prior_summary_cumulative}}
\centerline{\includegraphics[width=0.99\textwidth]{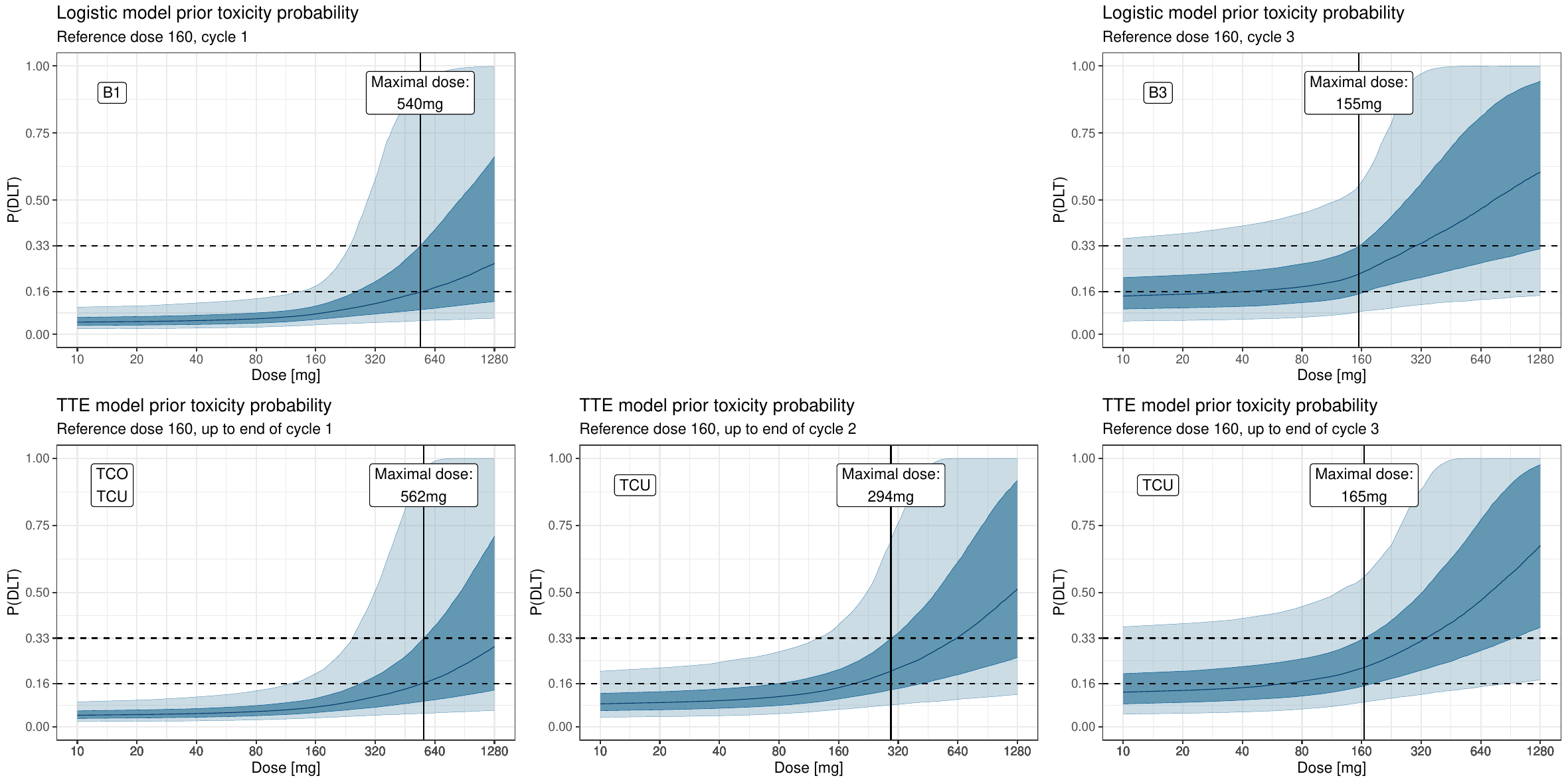}}
\end{subfigure}
\caption{Prior summary as a function of daily dose with a target toxicity band of 16\% to 33\%\label{fig:prior_summary}. \subref{fig:prior_summary_conditional} Conditional DLT probability by treatment cycle. The top left panel displays the dose-DLT prior for the 1-cycle BLRM (B1), whereas the bottom panels display the dose-DLT prior for the time-to-event model that limits toxicity conditionally by treatment cycle (TCO). \subref{fig:prior_summary_cumulative} Cumulative DLT probability over treatment cycles. The panels in the top row display the dose-DLT prior for the 1-cycle BLRM (B1; left panel) and 3-cycle BLRM (B3; right panel), whereas the bottom panels display the dose-DLT prior for the time-to-event model that limits cumulative toxicity over treatment cycles (TCU).}
\end{figure}

It is important to note that the EWOC criterion merely forbids to
choose certain doses which are labeled as too toxic and are also
referred to as overdose (here $\pi > 33\%$). Thereby, the EWOC defines a set of eligible doses to choose from whenever a decision is made during a trial conduct for the dose of a new patient cohort. Thus, to actually choose a dose to move forward with in a trial, one often uses other data than DLT events. A possible toxicity-related measure to guide dose escalation is the target probability for DLTs, which is the probability for a given dose to have a DLT rate in the range of $16$\%-$33$\% (also indicated in the prior dose-DLT summary in Figure~\ref{fig:prior_summary}). The dose with the highest target probability often coincides with the largest admissible dose, but this is not necessarily the case. Whenever the probability of DLT is less than $16$\%, it is classified as underdose in the following.

\subsection{Setup of models}
In the following we study a time-to-event and a logistic model with
either a focus on safety of a given cycle of a therapy (conditional on no previous events) or the entire therapy (cumulative events over all cycles). This results in four different models, which we here setup in an a consistent manner with respect to the prior. The reference dose $\tilde{d}$ plays a key role in defining the prior of the model. When the dose $d$ is equal to the reference dose $\tilde{d}$, then the slope term vanishes such that the mean of the intercept determines the prior DLT rate. 

While the reference dose $\tilde{d}$ concept suffices for defining the prior of the time-invariant logistic models, we suggest to introduce for the time-to-event models in addition the definition of a reference time-point $\tilde{t}$. This allows setting the prior for the mean of the intercept consistently with the logistic models by
using as reference time-point the end of a cycle $\tilde{j}$. As the
probability for an event within the time-interval $t$ is $\pi(t) =
1-S(t)$, it follows that for the defined reference
$\mbox{cloglog}(\pi(\tilde{t})) = \alpha + \log(\tilde{j}\,\Delta)$
holds. Accordingly, for the logistic models it holds that for the
reference $\mbox{logit}(\pi) = \alpha$. Hence, we set for the
time-to-event models the prior mean for $\alpha$ to
$\mbox{cloglog}(\tilde{p}) - \log(\tilde{j} \, \Delta)$ and for the
logistic models equivalently to $\mbox{logit}(\tilde{p})$. As
distribution a normal density is used with the standard deviation set
to unity allowing for substantial deviation from the prior mean
(recall that the parameter scale is $\mbox{logit}$ or the similar
$\mbox{cloglog}$ scale). For the prior on the slope parameter a
log-normal distribution is used which constrains the slope to positive values ensuring a conservative relationship of higher drug amounts leading to higher DLT rates, as well as ensuring that asymptotically for very high doses, the DLT probability goes to 1. We also assume no drug-drug interactions, i.e., the two components independently give rise to DLT events.

\subsubsection{Choice of prior}
For the time-to-event models, we choose the following intercepts for the two treatments: 
\begin{align*}
    \alpha_1 & \sim \textrm{Normal}\left(\mathrm{cloglog}(0.09), 1\right) \\
    \alpha_2 & \sim \textrm{Normal}\left(\mathrm{cloglog}(0.11), 0.5\right),
\end{align*}
with a reference time point of $\tilde{t} = 3 \cdot 6 \cdot 7 \mathrm{
  days}$, reflecting that we expect both treatments to be relatively
safe, with the prior for the background treatment to be slightly more certain, and the experimental combination partner treatment expected to be slightly safer, but with higher uncertainty. We also allow for a monotonic effect on the DLT probability in each cycle to account for differences in DLT rates between cycles. Here, we center this effect on 0, where the full effect over three cycles is $(J-1)\gamma_2$, and choose a minimally-informative prior for how much of the effect is observed at the end of cycle 1 and 2:
\begin{align*}
\gamma_2 & \sim N\left(0, 0.5\right) \\
\xi_2 & \sim \mathrm{Dirichlet}(1, 1).
\end{align*}
The slope parameter is constrained to be positive by assigning a prior to the $\log$ transformed slope $\log(\beta_1)$ for which we use a normal prior 
\begin{align*}
    \log(\beta_1) \sim \N(0, \log(4)/1.96).
\end{align*}
Since the prior is set on the logarithmic scale, it implies a wide
range of slopes once exponentiated and is weakly informative in the
sense of encoding the scale of the slope parameter to vary around
unity, with 95\% of the slopes between 1/4 and 4. Furthermore,
the complementary log-log and the logistic transform are relatively
similar to one another, such that this choice is very similar to the
slope prior of the BLRM, making it a well-tested choice.

For the one-cycle and three-cycles BLRMs, we matched the priors to the
cumulative DLT probability of the time-to-event model as closely as
possible (for one or three cycles, respectively, see
figure~\ref{fig:prior_summary}), yielding the following prior for the
one-cycle BLRM:
\begin{align*}
\alpha_1 & \sim \mathrm{Normal}(\mathrm{logit}(0.03), 1) \\
\alpha_2 & \sim \mathrm{Normal}(\mathrm{logit}(0.04), 0.5) \\
\log(\beta_1) & \sim \N(0, 0.9),
\end{align*}
and for the three-cycle BLRM, respectively:
\begin{align*}
\alpha_1 & \sim \mathrm{Normal}(\mathrm{logit}(0.09), 1.3) \\
\alpha_2 & \sim \mathrm{Normal}(\mathrm{logit}(0.12), 0.7) \\
\log(\beta_1) & \sim \N(0, 1.2),
\end{align*}

\subsection{Scenarios}
\begin{figure}[bh]
\centering
\begin{subfigure}{\textwidth}
\caption{\label{fig:scenarios_conditional}}
\centering
\includegraphics[width=0.9\textwidth]{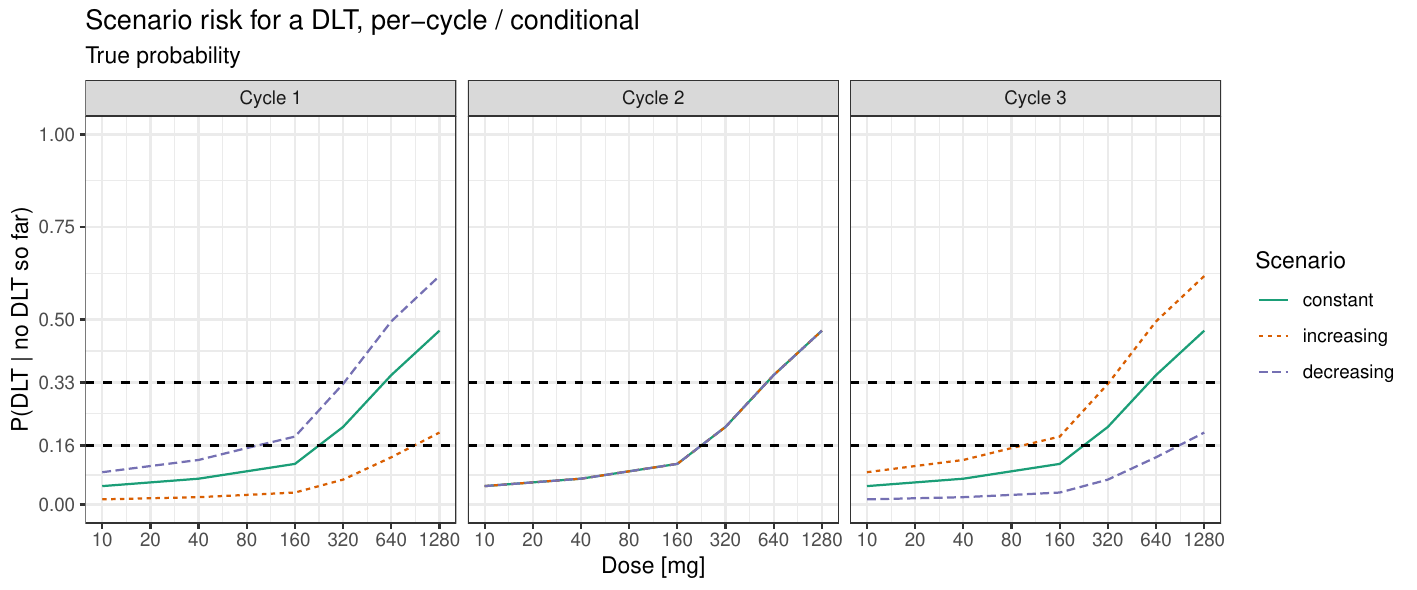}
\end{subfigure}
\begin{subfigure}{\textwidth}
\caption{\label{fig:scenarios_cumulative}}
\centering
\includegraphics[width=0.9\textwidth]{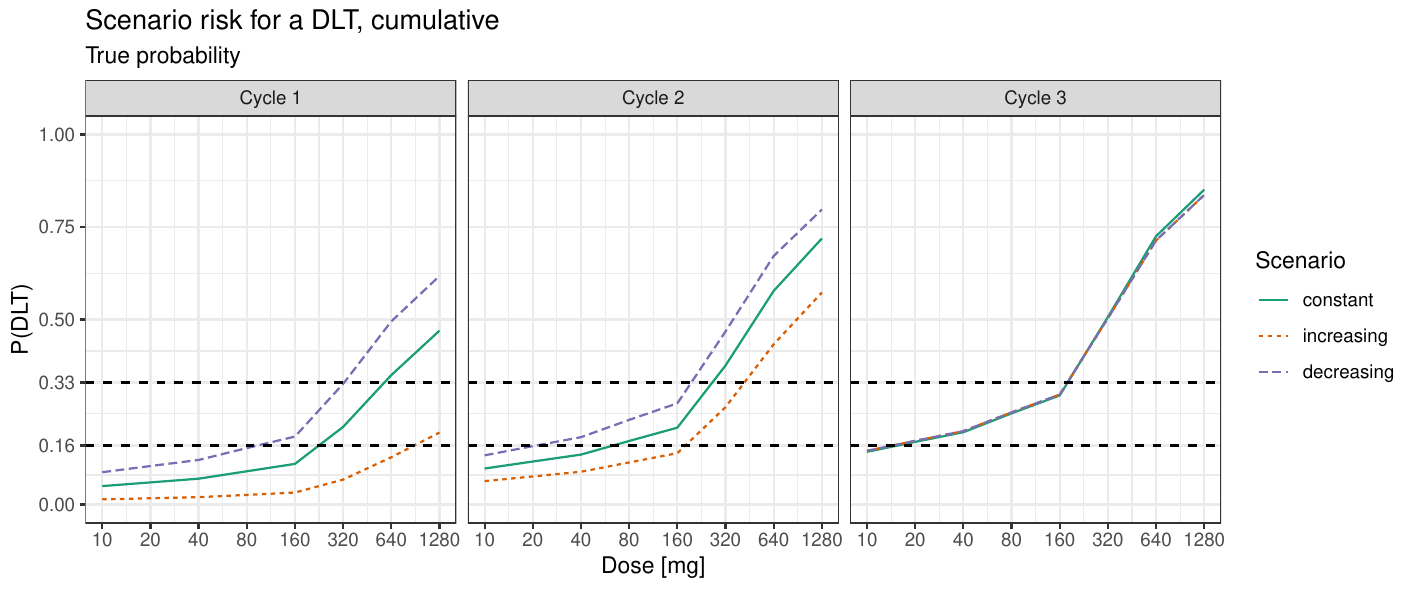}
\end{subfigure}
\caption{Summary of the ground truth scenarios (constant/increasing/decreasing conditional DLT probability over treatment cycles) and their (\subref{fig:scenarios_conditional}) conditional DLT probability in each treatment cycle, respectively, as well as their (\subref{fig:scenarios_cumulative}) cumulative DLT probabilities over treatment cycles.\label{fig:scenario_summary}}
\end{figure}
The experimental therapy for which the MTD is to be determined is
supposed to be developed for a dosing schedule to be given for three
cycles as outlined in the motivating case study in
Section~\ref{Case_Study}. Different toxicity scenarios and dropout scenarios are investigated. 

We chose our toxicity scenarios to have constant, increasing or decreasing conditional DLT probability over time for the three cycles respectively. This allows us to investigate the following toxicity scenarios:
\begin{description}
    \item[Constant:] a hypothetical drug has a memory-less
      DLT-generating process. This corresponds to the case where the
      dose-DLT probability curve is the same across all cycles
      conditionally on entering it.
    \item[Increasing:] a hypothetical drug (e.g., due to drug
      accumulation or damage) has a higher DLT probability for a given
      dose at later cycles. We note that this is a worst case scenario
      for 1-cycle methods, as all doses in cycle 1 are considered
      safe.
    \item [Decreasing]: a hypothetical drug presents with early
      DLTs (e.g., tumor lysis syndrome after start of cytotoxic
      chemotherapy). In this scenario, cycle 1 methods may prevent
      recommending too high doses, but it will nevertheless be
      relevant to see how time-to-event models are dealing with
      such profiles.
\end{description}
These scenarios are chosen in a manner that all of the dose-DLT-curves coincide in terms of conditional DLT probability of cycle 2, and have the same dose-DLT-curves in terms of cumulative toxicity over three cycles (see the visual summary provided in
Figure~\ref{fig:scenario_summary}).

Regarding dropout, in addition to above toxicity scenarios, we varied the probability of
dropout from 0\% through 33\% to 55\% across three treatment cycles
(constant over time as a memoryless process, see Table~\ref{tbl:dropout_scenarios}). The
following three dropout scenarios were investigated:

\begin{description}
    \item[Constant] (0\%, 33\%, 55\%) dropout rate over the entire
      dosing range.
    \item[Decreasing] dropout rate with increasing dose (step 
     function from 55\% at the lower half of the dose range to 0\% at the upper half), corresponding to, e.g., informative
     dropout for lack of efficacy.
    \item[Increasing] dropout rate with increasing dose (step function from 0\% at the lower half of the dose range to 55\% at the upper half), corresponding to, e.g., informative
      dropout due to tolerability issues at higher doses not
      manifesting in DLTs.
\end{description}

\begin{table*}[h]
\centering
\caption{Overview of non-informative / constant and informative / non-constant dropout scenarios across the range of pre-specified doses.  Reported rates for a memory-less process over 3 treatment cycles in the absence of competing DLT events. \label{tbl:dropout_scenarios}}
\begin{tabular}{rrr}\toprule
& \multicolumn{2}{c}{\makecell{Dropout rate over three \\ cycles in dose range}} 
\\\cmidrule(lr){2-3}
\makecell{~\\Dropout scenario}          & \makecell{Lower half: \\ 10 - 80 mg} & \makecell{Upper half:\\160 - 1280 mg} \\\midrule
Constant 0\%  & $0\%$ & $0\%$ \\
Constant 33\% & $33\%$ & $33\%$ \\
Constant 55\%  & $55\%$ & $55\%$ \\
Decreasing & $55\%$ & $0\%$ \\ 
Increasing & $0\%$ & $55\%$ \\
\bottomrule
\end{tabular}
\end{table*}

\subsection{Simulation study design}
The performance of the following 4 models was assessed in a simulation study targeting to determine the therapy-specific MTD over a 3-cycle observation period defined as follows:
\begin{description}
    \item[B1] BLRM based on a single cycle DLT observation period
    \item[B3] BLRM based on 3-cycle DLT observation period
    \item[TCO] Time-to-event based on conditional of no previous event
    \item[TCU] Time-to-event based on cumulative events over all cycles
\end{description}

The first cohort of a trial starts at the second-lowest dose level
(dose level of 20 mg), which allows the trial to proceed with the
lowest dose level in case the starting dose is found to be relatively toxic. For simplicity, every cohort includes 3 patients. The escalation process continues until the MTD is declared whenever a
given dose level fulfills the constraints that at least 6 patients
were treated at the dose level in question, the next dose level enrolled would be unchanged, and that the dose is with
more than $50$\% in the target toxicity range of 16\%-33\% toxicity
rate. Alternatively, at least 12 patients were treated at the given
dose being consistent with EWOC. Patient accrual was modelled as
a memoryless process with an accrual rate of 1 patient every 10 days
on average. For each of the scenarios, 1000 simulated trial replications were run. 

For the B1 and the time-to-event (TCO and TCU) methods, a decision for the next dose was made after the first cycle was completed by all
enrolled patients. For B3, all three cycles must have been
completed. In case all patients in a cohort dropped out until the end
of cycle 1 (or cycle 3 in the case of B3), another cohort was enrolled at the same dose level.

The MTD is declared once the data from cycle 3 of the last patient
enrolled in a trial is available (or that patient has dropped out earlier), i.e., once the data acquisition process is completed. Alternatively, if all the pre-specified dose levels do not fulfill EWOC, the trial is aborted without MTD declaration. The maximum sample size at which the simulation would stop in case the MTD was not found yet was set to 60 patients. None of the replications in any of the scenarios reached this threshold, that is, none of the trials were stopped due to reaching the maximum sample size. For each trial replication that declared MTD, we then assess whether it falls into the overdosing, target or underdosing intervals for the given scenario.

The simulation code was implemented in R, where  the time-to-event models were implemented using the brms package\cite{Buerkner2017} and the BLRMs using the OncoBayes2 package \cite{OncoBayes2}. Simulations were run in parallel on 200 cores using the clustermq package \cite{Schubert2019}.

\section{Results}\label{Results}
The results of the simulation study are summarized below in terms of
three metrics: (i) correctness of the MTD declarations as relevant to the
drug development program (treatment-specific MTD), (ii) probability to treat patients with doses of respective risk classes illustrating the risk to patients within each trial and (iii) the trial duration and number of recruited patients (sample size) reflecting operational properties of a given design.

\begin{landscape}
\begin{figure}[!h]
\centering
    \centerline{\includegraphics[width=1.25\textwidth]{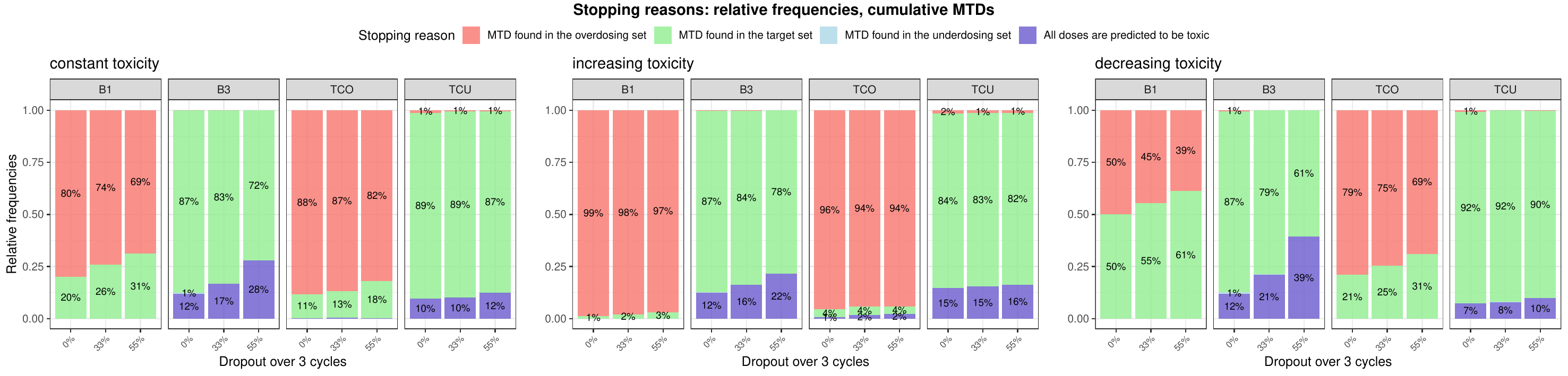}} 
    \caption{Summary of MTD declaration probabilities by scenario, method and (non-informative) dropout. Each bar represents 1000 trial replications\label{fig:MTD_cumulative_frequencies}.}
\end{figure} 
\begin{figure}[!h]
    \centerline{\includegraphics[width=1.25\textwidth]{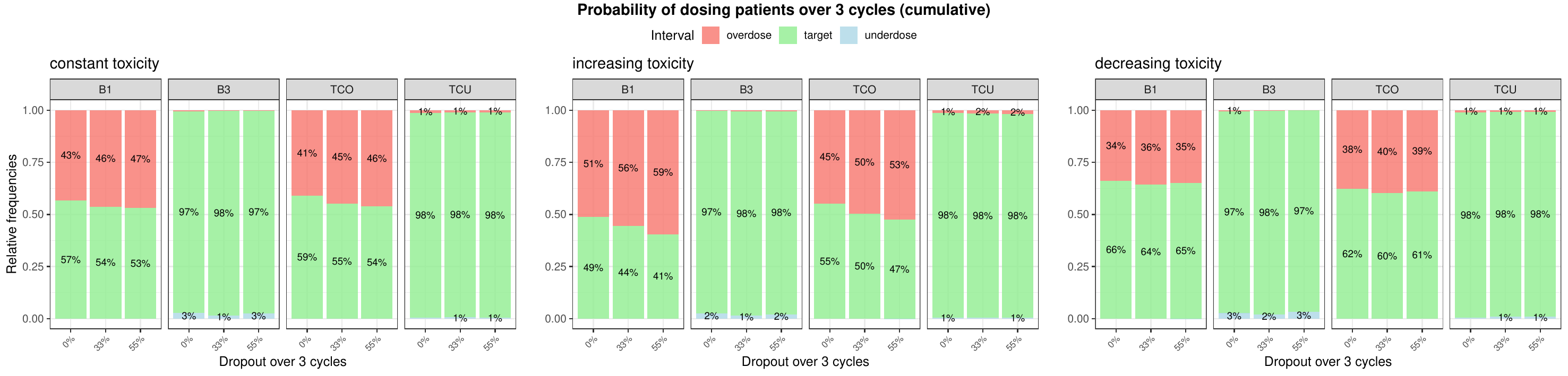}} 
    \caption{Summary of patient allocation probabilities by scenario, method and (non-informative) dropout. Each bar represents the sum of allocated patients across 1000 trial replications\label{fig:patient_cumulative_allocation}.}
\end{figure}
\end{landscape}

\subsection{Accuracy of MTD declaration}
The accuracy of MTD declaration is summarized in
Figure~\ref{fig:MTD_cumulative_frequencies}. The 1-cycle and
per-cycle-models (B1 and TCO) almost always declare MTD, but most
likely, the declared treatment-specific MTD is too toxic. This holds across all of the scenarios of increasing, constant and decreasing conditional toxicity probability over treatment cycles, though is the most pronounced in the increasing toxicity case (where many DLT events occur in cycles 2 and 3). Methods that assess toxicity over three cycles (B3 and TCU) perform much better in all toxicity
scenarios. While the mean probability to select a treatment-specific MTD in the target range is only $20\% \pm 1\%$ (mean, Monte Carlo Standard Error from here on $\leq 1\%$ unless otherwise stated) and $12\%$ for the constant toxicity / no dropout scenario for the B1 and TCO models, the probability is $87\%$ and $89\%$ for the B3 and TCU models, respectively. 

As expected, for the increasing toxicity scenario the outcome is even worse. The probability to identify the MTD in the target range is only $1\%$ for B1 and $4\%$ for TCO. The probability to determine the treatment-specific MTD in the target range is $87\%$ and $84\%$ for the B3 and TCU models. It
is almost certain to select a treatment-specific MTD in the overdosing range when using the B1 and TCO model under the constant or increasing toxicity scenario. 

We may anticipate that both B1 and TCO have a better accuracy under the decreasing toxicity profile, where more DLT events happen in cycle 1. This is the case for the B1 model which identifies now an MTD in the target range with a probability of $50\% \pm 2\%$. However, the TCO model picks an MTD in the overdose range with a probability of $21\%$. The B3 and the TCU model are also performing better under the decreasing toxicity scenario with picking the treatment-specific MTD in the target range with probability of $87\%$ and $92\%$, respectively.

Across all scenarios, the B3 and TCU methods stop some trials for toxicity (when all doses are predicted to be toxic). This becomes more pronounced when non-informative dropout is introduced, especially for the B3 method in the decreasing (early) toxicity and constant toxicity scenarios, where the fraction of stopped trials increases from $12\%$ to $39\% \pm 2\%$ and $12\%$ to $28\%$, respectively. For the TCU method, this is less pronounced, with an increase from $7\%$ to $10\%$ and $10\%$ to $12\%$, respectively. The TCU method stops the most trials under the increasing toxicity scenario, ranging from $15\%$ to $16\%$ with increasing dropout rate, whereas B3 ranges from $12\%$ to $22\%$.

When non-informative dropout is introduced, we see conservative
behaviour of all models in terms of recommending lower doses with increasing dropout rate. In particular, B3 gets very conservative in the decreasing toxicity scenario where DLTs mostly occur in the first treatment cycle. 

When dropout is informative: Under the constant toxicity scenario MTD declarations as summarized in Figure~\ref{fig:MTD_cumulative_frequencies_info} with decreasing dropout with dose, the outcome of the B1 and TCO models is similar to not having dropout present. When dropout increases with dose, the outcome is similar to having 55\% dropout at all doses. For B3 and TCU models, the reverse is the case: dropout decreasing with dose yields results close to having 55\% dropout at all doses, whereas dropout increasing with dose is close to the no-dropout scenario. In summary, the dropout probability in the dose region where the MTD is declared seems to be most relevant.

\begin{figure}[t]
\centerline{\includegraphics[width=0.35\textwidth]{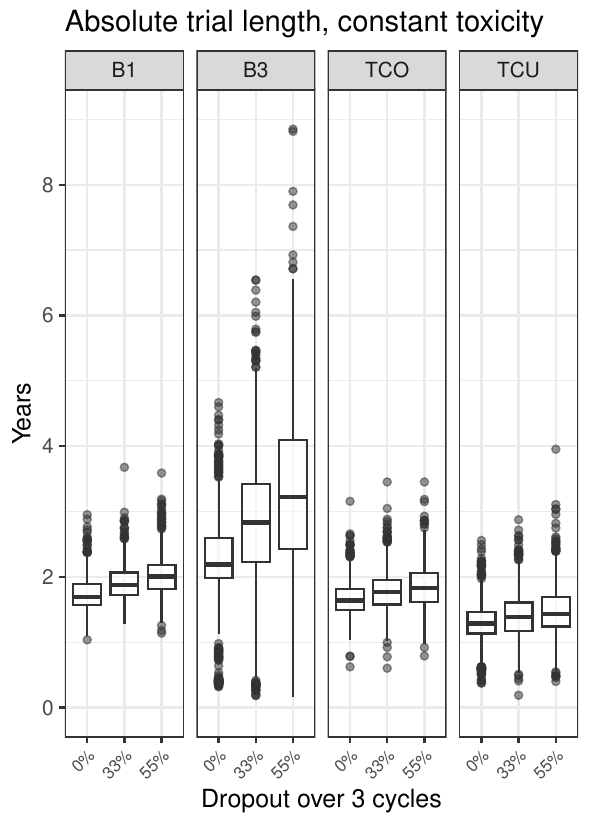}} 
\caption{Trial length from first patient enrollment to end of data acquisition in the constant toxicity scenario\label{fig:trial_lengths_main}.}
\end{figure}

\begin{figure}[t]
\centerline{\includegraphics[width=\textwidth]{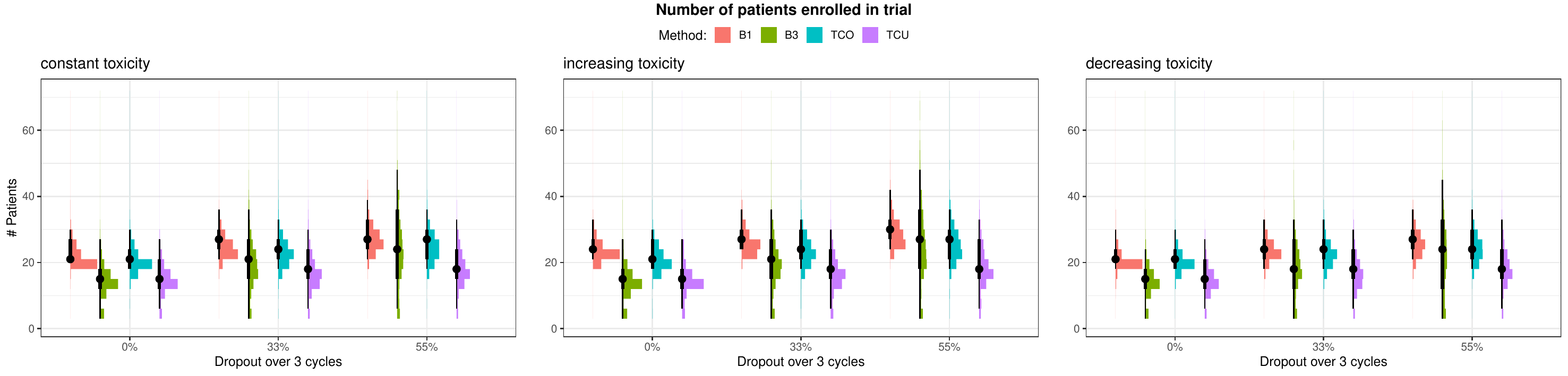}} 
\caption{Number of patients enrolled per trial\label{fig:trial_npatients_main}.}
\end{figure}

\subsection{Risks to patients in the trials}
B1 and TCO methods exhibit systematically higher probability of being exposed to doses which are too high over three cycles than the methods that control toxicity over three cycles (B3 and TCU). These results are summarized in Figure~\ref{fig:patient_cumulative_allocation}. This finding holds across all of the scenarios of increasing, constant and decreasing conditional toxicity probability over treatment cycles, although it is most pronounced in the increasing toxicity scenario. For the B1 and the TCO model we observe also some increasing risk of being exposed to a dose which is too high with increasing non-informative dropout rate. For the B3 and TCU models there is no impact of the dropout rate on the risk of being exposed to doses which are too high. This finding holds also if dropout is informative, regardless whether de- or increasing with dose (Figure~\ref{fig:patient_cumulative_allocation_info}). 

\subsection{Trial conduct: sample size and trial length}
As pointed out in Section~\ref{Introduction}, sample size and trial length are critical factors for drug development. The absolute trial lengths in terms of time elapsed between enrolling the first patient and declaring MTD or stopping the trial for toxicity were markedly different (Figure~\ref{fig:trial_lengths_main} for the constant toxicity scenario and Figure~\ref{fig:incr_decr_trial_lengths} for the increasing/decreasing scenarios). In particular, the B3 approach takes most time  and the TCU approach is fastest in all toxicity scenarios. In general, all methods get slower with increasing dropout rate. This increase can be substantial for the B3 approach while the impact of dropout is rather small for the other models. We observed extreme cases of experiments taking 8+ years for B3 for the 55\% dropout scenario. In addition we note that the variability of the trial length is small and similar for the models other than B3. Variability for the B3 model is substantial, especially under dropout. We see little impact of the toxicity scenario on the trial length regardless of the approach. There is also no relevant difference on the trial lengths if drop-out is informative or not (Figure~\ref{fig:trial_lengths_info}). 

For all toxicity scenarios, the TCU model required on average the fewest patients (as measured in terms of patients \emph{enrolled}, regardless of dropout) to determine the MTD (Figure~\ref{fig:trial_npatients_main}). For the constant toxicity scenario without dropout, the sample size was comparable for B1 and TCO, as well as B3 and TCU, respectively (mean $\pm$ MCSE for B1: $n=23.1 \pm 0.1$, B3: $n=15.5 \pm 0.2$, TCO: $n=21.9 \pm 0.1$, TCU: $n=16.2 \pm 0.2$). As expected, the sample size increased with increasing dropout rate for all toxicity scenarios for all models and scenarios, with the sample size more rapidly increasing in the methods using binary rather than time-to-event data, and the TCU method requiring the least samples under substantial dropout of 55\% over three cycles.

\subsection{Summary of simulation study results}
As illustrated by the simulation study, the models are addressing different clinical questions. It is critical that the model is aligned with the underlying clinical question of interest: for any setting where drugs are intended to be given for a longer period of time, 1-cycle methods are not recommended. This holds even more when the dose-toxicity relationship changes over time. Multi-cycle approaches address the more relevant clinical question regarding a treatment-specific MTD, which is especially relevant for targeted or immune compounds taken for prolonged periods of time. 

Overall, TCO and B1 are likely to recommend MTD which is too high and likely to expose more patients to overdose compared to TCU and B3. We note that TCO is not a good choice: There is a high probability that the treatment-specific MTD is too high in particular in the presence of dropout, regardless of whether informative or not. B1 might be a model of interest if we believe that there is only early toxicity (such as for the cytotoxic treatments the method was originally designed for).

TCU is highly recommended: This model provides most efficient characteristics with respect to time and patient numbers, it is not sensitive to both drop-out (informative or not) and in- or decreasing toxicity. It is very likely that this model determines the MTD in the target range (except for extreme scenarios).

B3 presents a poorer performance compared to TCU, especially under dropout. There is higher risk to stop a trial too early for toxicity compared to TCU and is therefore not recommended.

\section{Discussion}\label{Discussion}

With the common shift in the oncology treatment paradigm of modern targeted or immuno-therapies to transform an acute disease into a chronic condition, dose-finding requires a sensitive methodology to determine a dose level that can be tolerated over a longer treatment period. Classical BLRM-like models which are based on a 1-cycle DLT observation period are no longer answering the relevant clinical question of interest. In our simulation study we showed that the 1-cycle BLRM and TCO models generally recommend a dose which is too high and also expose more patients to toxic doses compared to the 3-cycle models like the TCU and the B3. TCU in particular seems to be a robust choice when it comes to considering the dose-toxicity relationship over the course of a treatment with partial data and under dropout. 

For the TCO method, for the decreasing toxicity / early toxicity case, considering the toxicity of the other cycles (which is lower) makes dosing recommendations in cycle 1 overly toxic more often than for B1 -- this likely also depends on how strong our prior on similar dose-response in different cycles is, i.e., the monotonic term. If one left this term away, the per-cycle toxicity would be constant, i.e., if there is different timing of DLTs, this gets smeared out over all cycles, and each individual cycle will likely look safe to the model (even if it really is not).

In common approaches, decision-making on doses is bound to predefined criteria when a certain number of patients have completed the DLT observation period or when unexpected events occurred, like the first 2 patients of a cohort having experienced a DLT. The time-to-DLT approach allows more flexible timing of decision making: Whenever a patient completed a cycle-defined DLT observation period, the model can be updated for decision making. Continuous monitoring of patients and dose assessments is possible. However, since the likelihood of a DLT might change over time, a final analysis should be performed whenever all patients of a dose cohort have completed all cycles of interest. Assessing the dose-toxicity relationship based on multiple cycle changes the operational requirements. It is important to monitor patients throughout the follow-up period and to assess toxicities at latest at the end of each cycle. 

Trialists need to define stopping criteria in the study protocol which require not only to be based on the number of evaluable patients but also regarding the minimum information in terms of patient-cycles. Simulations including potential informative drop-out considering different dose-toxicity profiles are useful to assess the stopping criteria and the required information. Since we are interested in patient-cycles, minimum exposure criteria would need to be defined at the patient-cycle level. Patient information might be interpretable for the first cycles only. 

Dropout will be a more common phenomenon when it comes to multi-cycle assessments rather than 1-cycle assessments, as well as in populations with rapidly progressing disease. Special attention is required to document the reason why follow-up could not be completed if the reason is different from experiencing a DLT. Reasons such as dropout due to disease progression or due to non-DLT AEs might be informative for dose-finding and their impact could be assessed in supplementary and/or sensitivity analyses.

Study duration is a critical metric in drug development and approaches requiring more than 1 cycle of observation period will take more time. In the simulation study, the B3 model takes substantially more time than all other methods (Figure \ref{fig:trial_lengths_main}). Since dose assessments are only done if all patients of a cohort have completed all scheduled cycles, the duration is substantially longer compared to other models in particular in the presence of dropout. Another phenomenon is the efficiency of the B3 model in terms of patient numbers: All other approaches require more patients. However, this holds only in the case of no dropout. If there is dropout, the B3 model has a high variability in the number of required patients. Dropout and DLT are competing events. The higher the probability of DLT, patients are more likely to stop for DLT than for dropping out.

Informative dropout seems to have the biggest effect when the dropout is high around where the MTD would be declared by the method (B1: drop-out increasing with dose is similar to non-informative dropout of 55\%, B3: decreasing with dose is similar to non-informative 55\%).
Note that the impact of dropout might be different in scenarios with more pronounced changes in the dose-toxicity relationship over time than we investigated in the scenarios, in settings where the DLT observation period is even longer, or in populations with rapidly progressing disease. 

We implemented the time to DLT model considering the typical DLTs as event of interest. However, the common DLT definition which includes quite severe side effects might not be sensitive enough to guide dose decisions for a longer-term treatment. However, the model can also use less severe events than DLTs like lower grade AEs or a predefined worsening of a PRO scale, e.g. CTC-PRO.

Although we call our model a time-to-event approach, it is based on discrete patient-cycle intervals. This might come with some loss of information when there is a difference in the early onset of toxicity. An aspect of further development is to use the actual time instead of discrete cycle intervals. We also acknowledge that delays in starting the next cycle due to toxicities are common and thus, prolonged cycle lengths, although not meeting DLT criteria, might also be informative for dose finding. 

\subsection*{Author contributions}
\textbf{Lukas A. Widmer}: Conceptualization, Methodology, Formal Analysis, Software, Visualization, Writing - Original Draft Preparation. \textbf{Sebastian Weber}: Conceptualization, Methodology, Formal Analysis, Software, Visualization, Writing - Original Draft Preparation. \textbf{Yunnan Xu}: Conceptualization, Writing - Original Draft Preparation. \textbf{Hans-Jochen Weber}: Conceptualization, Supervision, Writing - Original Draft Preparation.

\subsection*{Financial disclosure}

Lukas A. Widmer, Sebastian Weber and Hans-Jochen Weber are employed by Novartis Pharma AG and own stocks in the company. Yunnan Xu is employed by Novartis Pharmaceuticals Corporation and owns stocks in the company.

\subsection*{Conflict of interest}

The authors declare no potential conflict of interests.

\section*{Supporting information}

The following supporting information is available as part of the online article:

\subsection*{Code}
\textbf{Git repository} at \url{https://github.com/luwidmer/tte-manuscript}, containing the the R source code that sets up and runs the simulation study, as well as produces the figures in this manuscript.

\noindent

\appendix
\supplementarysection

\begin{figure}
    \centering
    \centerline{\includegraphics[width=.9\textwidth]{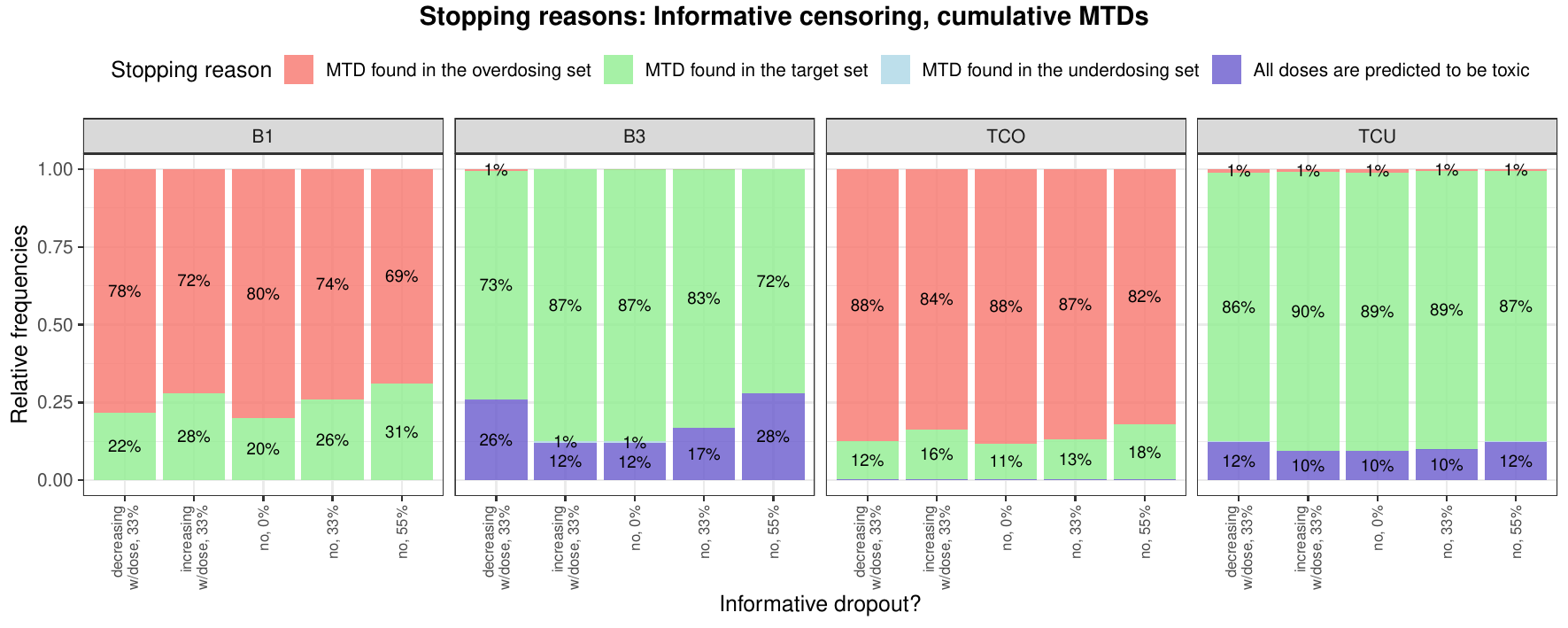}} 
    \caption{Summary of MTD declaration probabilities for the constant conditional toxicity scenario, by method and for non-informative and informative dropout rate scenarios. Each bar represents 1000 trial replications\label{fig:MTD_cumulative_frequencies_info}.}
\end{figure} 
\begin{figure}
    \centering
    \centerline{\includegraphics[width=.9\textwidth]{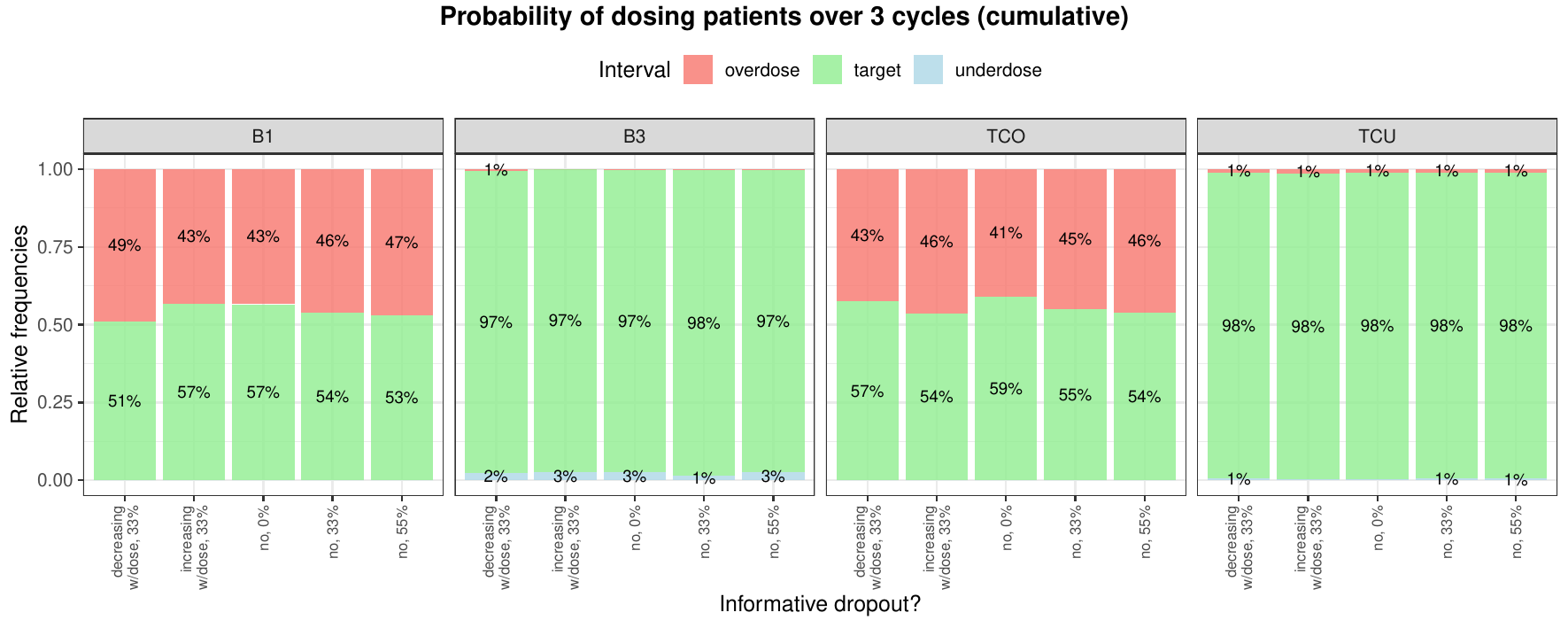}} 
    \caption{Summary of patient allocation probabilities for the constant conditional toxicity scenario, by method and for non-informative and informative dropout rate scenarios. Each bar represents the sum of allocated patients across 1000 trial replications\label{fig:patient_cumulative_allocation_info}.}
\end{figure}

\begin{figure}
\centerline{\includegraphics[width=0.9\textwidth]{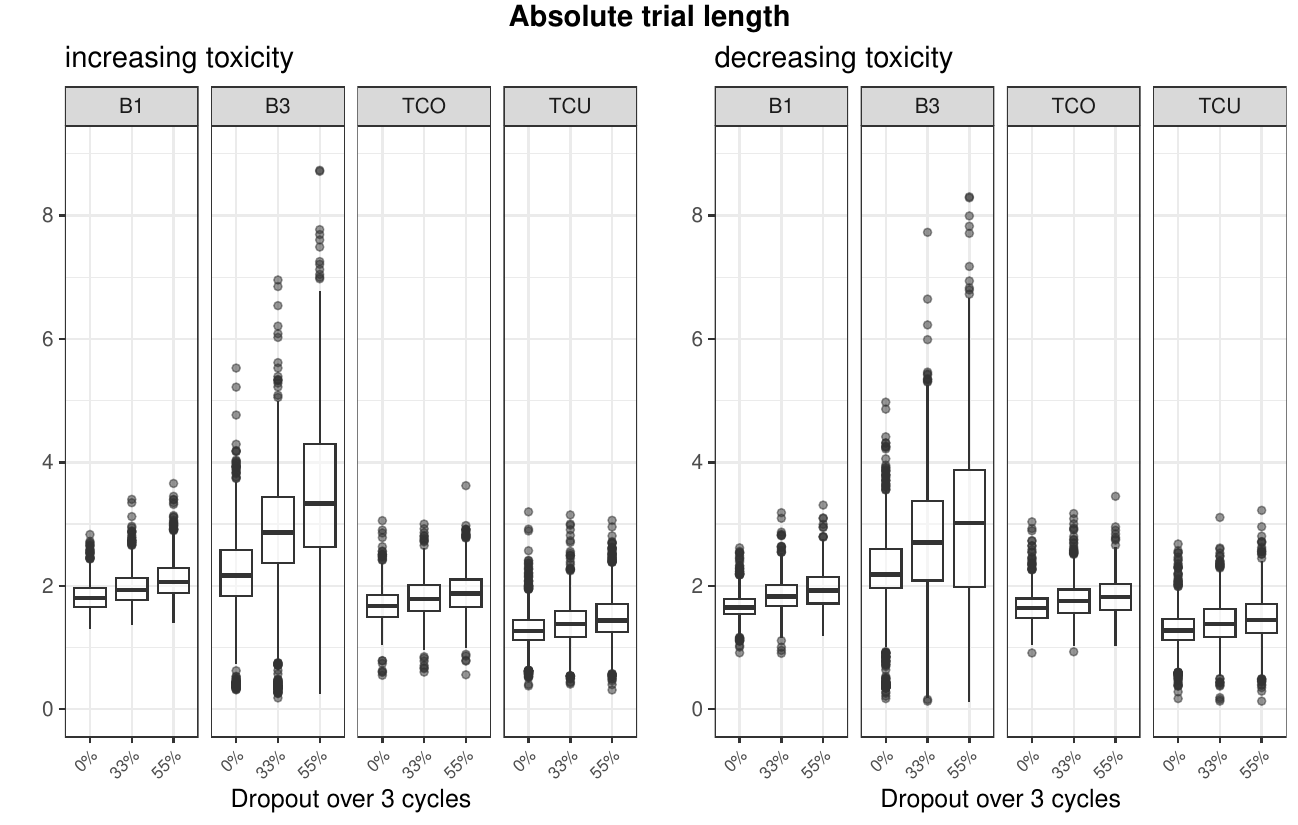}} 
\caption{Trial lengths for increasing and decreasing toxicity scenarios \label{fig:incr_decr_trial_lengths}}
\end{figure}

\begin{figure}
\centerline{\includegraphics[width=0.8\textwidth]{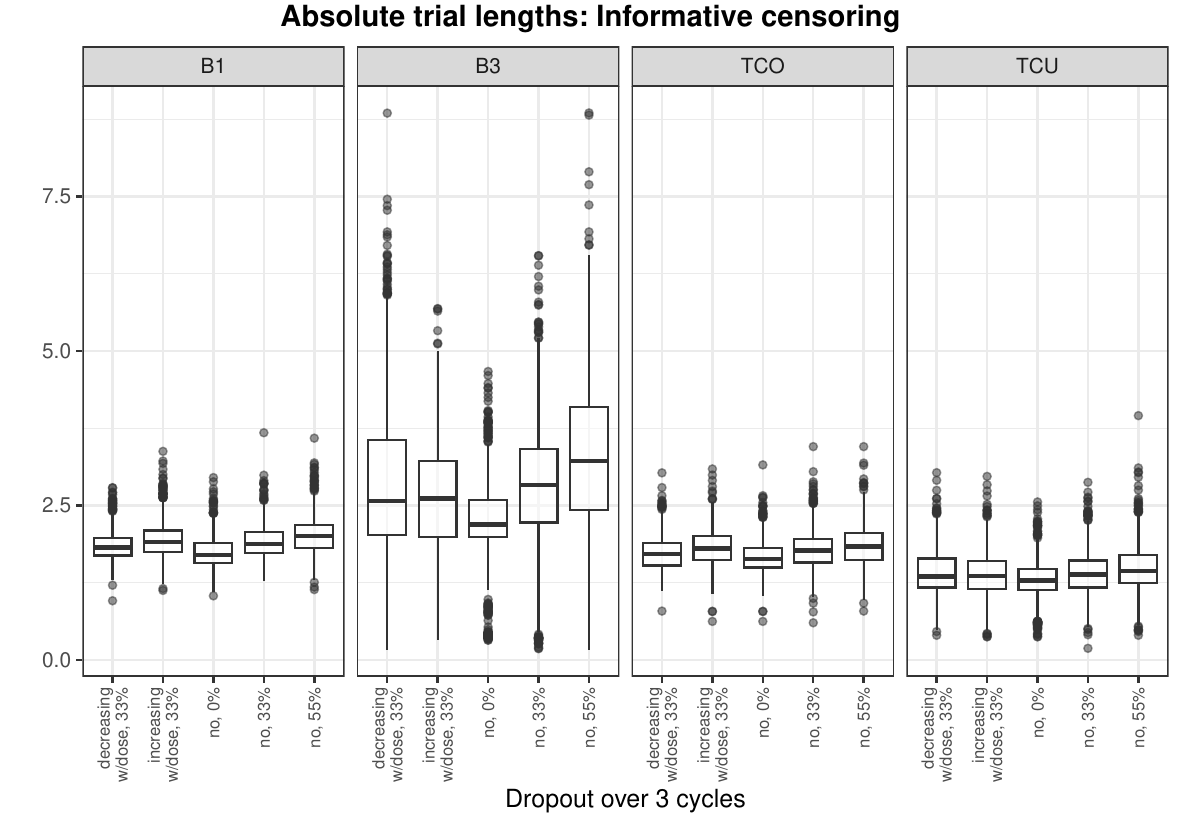}} 
\caption{Trial lengths for informative censoring scenarios \label{fig:trial_lengths_info}}
\end{figure}

\clearpage

\bibliography{wileyNJD-AMA}%

\end{document}